RESEARCH　　　　　　　　　　　　　　　　　　　　　　　　　　　　　　　　　　　　　　　　Open Access

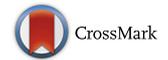

# Pretata: predicting TATA binding proteins with novel features and dimensionality reduction strategy

Quan Zou[1], Shixiang Wan[1,2], Ying Ju[3], Jijun Tang[1,4] and Xiangxiang Zeng[3*]

## Abstract

**Background:** It is necessary and essential to discovery protein function from the novel primary sequences. Wet lab experimental procedures are not only time-consuming, but also costly, so predicting protein structure and function reliably based only on amino acid sequence has significant value. TATA-binding protein (TBP) is a kind of DNA binding protein, which plays a key role in the transcription regulation. Our study proposed an automatic approach for identifying TATA-binding proteins efficiently, accurately, and conveniently. This method would guide for the special protein identification with computational intelligence strategies.

**Results:** Firstly, we proposed novel fingerprint features for TBP based on pseudo amino acid composition, physicochemical properties, and secondary structure. Secondly, hierarchical features dimensionality reduction strategies were employed to improve the performance furthermore. Currently, Pretata achieves 92.92% TATA-binding protein prediction accuracy, which is better than all other existing methods.

**Conclusions:** The experiments demonstrate that our method could greatly improve the prediction accuracy and speed, thus allowing large-scale NGS data prediction to be practical. A web server is developed to facilitate the other researchers, which can be accessed at http://server.malab.cn/preTata/.

## Background

TATA-binding protein (TBP) is a kind of special protein, which is essential and triggers important molecular function in the transcription process. It will bind to TATA box in the DNA sequence, and help in the DNA melting. TBP is also the important component of RNA polymerase [1]. TBP plays a key role in health and disease, specifically in the expression and regulation of genes. Thus, identifying TBP proteins is theoretically significant. Although TBP plays an important role in the regulation of gene expression, no studies have yet focused on the computational classification or prediction of TBP.

Several kinds of proteins have been distinguished from others with machine learning methods, including DNA-binding proteins [2], cytokines [3], enzymes [4], etc. Generally speaking, special protein identification faces three problems, including feature extraction from primary sequences, negative samples collection, and effective classifier with proper parameters tuning.

Feature extraction is the key process of various protein classification problems. The feature vectors sometimes are called as the fingerprints of the proteins. The common features include Chou's PseACC representation [5], K-mer and K-ship frequencies [6], Chen's 188D composition and physicochemical characteristics [7],

* Correspondence: xzeng@xmu.edu.cn
[3]School of Information Science and Engineering, Xiamen University, Xiamen, China
Full list of author information is available at the end of the article

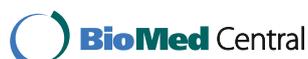

Wei's secondary structure features [8, 9], PSSM matrix features [10], etc. Some web servers were also developed for features extraction from protein primary sequence, including Pse-in-one [11], Protrweb [12], PseAAC [13], etc. Sometimes, feature selection or reduction techniques were also employed for protein classification [14], such as mRMR [15], t-SNE [16], MRMD [17].

Negative samples collection recently attracts the attention from bioinformatics and machine learning researchers, since low quality negative training set may cause the weak generalization ability and robustness [18–20]. Wei et al. improved the negative sample quality by updating the prediction model with misclassified negative samples, and applied the strategies on human microRNA identification [21]. Xu et al. updated the negative training set with the support vectors in SVM. They predicted cytokine-receptor interaction successfully with this method [22].

Proper classifier can help to improve the prediction performance. Support vector machine (SVM), k-nearest neighbor (k-NN), artificial neural network (ANN) [23], random forest (RF) [24] and ensemble learning [25, 26] are usually employed for special peptides identification. However, when we collected all available TBP and non-TBP primary sequences, it was realized that the training set is extremely imbalanced. When classifying and predicting proteins with imbalanced data, accuracy rates may be high, but resulting confusion matrices are unsatisfactory. Such classifiers easily over-fit, and a large number of negative sequences flood the small number of positive sequences, so the efficiency of the algorithm is dramatically reduced.

In this paper, we proposed an optimal undersampling model together with novel TBP sequence features. Both physicochemical properties and secondary structure prediction are selected to combine into 661 dimensions (661D) features in our method. Then secondary optimal dimensionality searching generates optimal accuracy, sensitivity, specificity, and dimensionality of the prediction.

## Methods
### Features based on composition and physicochemical properties of amino acids

Previous research has extracted protein feature information according to composition/position or physicochemical properties [27]. However, analyzing only either composition/position or physicochemical properties alone does not ensure that the process is comprehensive. Dubchak proposed a composition, transition, and distribution (CTD) feature model in which composition and physicochemical properties were used independently [28, 29]. Cai et al. developed the 188 dimension 188D feature extraction method, which combines amino acid compositions with physicochemical properties into a functional classification of a protein based on its primary sequence. This method involves eight types of physicochemical properties, namely, hydrophobicity, normalized van der Waals volume, polarity, polarizability, charge, surface tension, secondary structure, and solvent accessibility. The first 20 dimensions represent the proportions of the 20 kinds of amino acids in the sequence. Amino acids can be divided into three categories based on hydrophobicity: neutral, polar, and hydrophobic. The neutral group contains Gly, Ala, Ser, Thr, Pro, His, and Tyr. The polar group contains Arg, Lys, Glu, Asp, Gln, and Asn. The hydrophobic group contains Cys, Val, Leu, Ile, Met, Phe, and Trp [30].

The CTD model was employed to describe global information about the protein sequence. C represents the percentage of each type of hydrophobic amino acid in an amino acid sequence. T represents the frequency of one hydrophobic amino acid followed by another amino acid with different hydrophobic properties. D represents the first, 25%, 50%, 75%, and last position of the amino acids that satisfy certain properties in the sequence. Therefore, each sequence will produce 188 (20 + (21) × 8) values with eight kinds of physicochemical properties considered.

The 20 kinds of amino acids are denoted as $\{A_1, A_2, ..., A_{19}, A_{20}\}$, and the three hydrophobic group categories are denoted as [n, p, h].

In terms of the composition feature of the amino acids, the first 20 feature attributes can be given as

$$E_i = \text{Number of } A_i \text{ in sequence} \mid \text{Length of sequence} \times 100, \ (1 \leq i \leq 20)$$

Extracted features are organized according to the eight physicochemical properties. $D_i$ (i = n, p, h) represents amino acids with i hydrophobic properties. For each hydrophobic property, we have

$$C_i = \text{number of } D_i \text{ in sequence} \mid \text{length of sequence} \times 100, \ (i = n, p, h)$$

$$T_{ij} = \text{number of pairs like } D_i D_j \text{ or } D_j D_i \mid [(\text{length of sequence}) - 1] \times 100$$

where $i, j \in \{(i = n, j = p), (i = n, j = h), (i = p, j = h)\}$.

$$D_{ij} = P_j \text{th position of } D_i \mid \text{length of sequence} \times 100, \ (j = 0, 1, 2, 3, 4; i = n, p, h)$$

$$P_j = \begin{cases} 1, j = 0 \\ \lfloor N \mid 4 \times j \rfloor \end{cases}, (j = 1, 2, 3, 4; N = \text{number of } D_i \text{ in sequence})$$

Based on the above feature model, the 188D features of each protein sequence can be obtained.

### Features from secondary structure

Secondary structure features were proved to be efficient for representing proteins. They contributed on the protein fold pattern prediction. Here we try to find the well worked secondary structure features for TBP identification. The PSIPRED [31] protein structure prediction server (http://bioinf.cs.ucl.ac.uk/psipred/) allows users to submit a protein sequence, perform the prediction of their choice, and receive the results of that prediction both textually via e-mail and graphically via the web. We focused on PSIPRED in our study to improve protein type classification and prediction accuracy. PSIPRED employed artificial neural network and PSI-BLAST [32, 33] alignment results for protein secondary structure prediction, which was proved to get an average overall accuracy of 76.5%. Figure 1 gives an example of PSIPRED secondary structure prediction.

Then we viewed the predicted secondary structure as a sequence with 3-size-alphabet, including H($\alpha$-helix), E($\beta$-sheet), C($\gamma$-coil). Global and local features were extracted from the secondary structure sequences. The total of the secondary structure is 473D.

### Features dimensionality reduction

The composition, physicochemical and secondary structure features are combined into 611D high dimension feature vectors. We try to employ the feature dimensionality reduction strategy for delete the redundant and noise features. If two features are highly dependent on one another, their contribution toward distinguishing a target label would be reduced. So the higher the distance between features, the more independent those features become. In this work, we employed our previous work MRMD [17] for features dimension reduction. MRMD could rank all the features according their contributions to the label classification. It also considers the feature redundancy. Then the important features would be ranked on top.

To alleviate the curse of high dimensionality and reduce redundant features, our method uses MRMD to reduce the number of dimensions from 661 features, and searches for an optimal dimensionality based on secondary dimension searching. MRMD calculates the correlation between features and class standards using Pearson's correlation coefficient, and redundancy among features using a distance function. MRMD dimension reduction is simple and rapid, but can only produce results one by one, and increases the actual computation time greatly. Therefore, based on the above analyses, we developed Secondary-Dimension-Search-TATA-binding to find the optimal dimensionality with the best ACC, as shown in Fig. 2 and Algorithm 1.

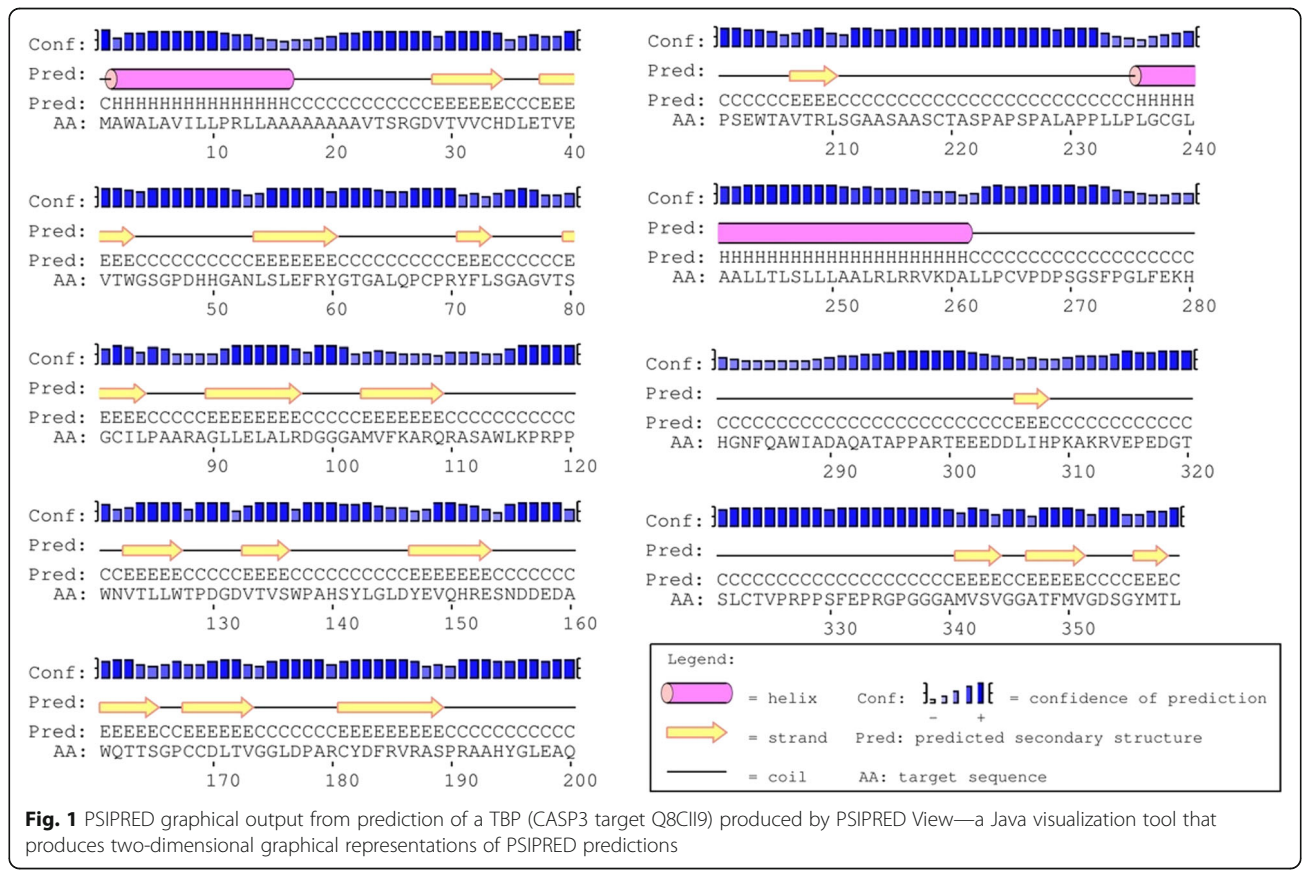

**Fig. 1** PSIPRED graphical output from prediction of a TBP (CASP3 target Q8CII9) produced by PSIPRED View—a Java visualization tool that produces two-dimensional graphical representations of PSIPRED predictions

Algorithm 1. Searching the best dimension based on MRMD

(1) Input: 661D (188D plus PSIPRED) dataset;

(2) Output: the best ACC, dimension $D$ and related results file;

(4) Select Initial Dimension, Multiple Iteration as *primary step*, Reduction Iteration as *secondary step*, and classifier $C$ (default is LIBSVM);

(5) Iterating as starting from Initial Dimension, and running 10-fold cross validation by $C$ to calculate ACC.

(6) If ACC is higher, go to the next search; if not, reduce dimension according to *secondary step* and repeat;

(7) Find the optimal dimension $D$ and the best ACC;

(8) Output the best ACC and D.

As described in Fig. 2 and Algorithm 1, searching the optimal dimension contains two sub-procedures: the coarse primary step, and the elaborate secondary step. The primary step aims to find large-scale dimension range as much as quickly. The secondary step is more elaborate searching, which aims to find specific small-scale dimension range to determine the final optimal accuracy, sensitivity and specificity. In the primary step, we define the initial dimension reasonably according to current dataset, and a tolerable dimension, which is also the lowest dimension. Based on this primary step, the dimensionality of sequences will become sequentially lower with MRMD analysis. After finding the best accuracy from all running results finally, the secondary step starts. In the secondary step, MRMD runs and scans all dimensions according to the secondary step sequentially to calculate the best accuracy, which likes the primary step.

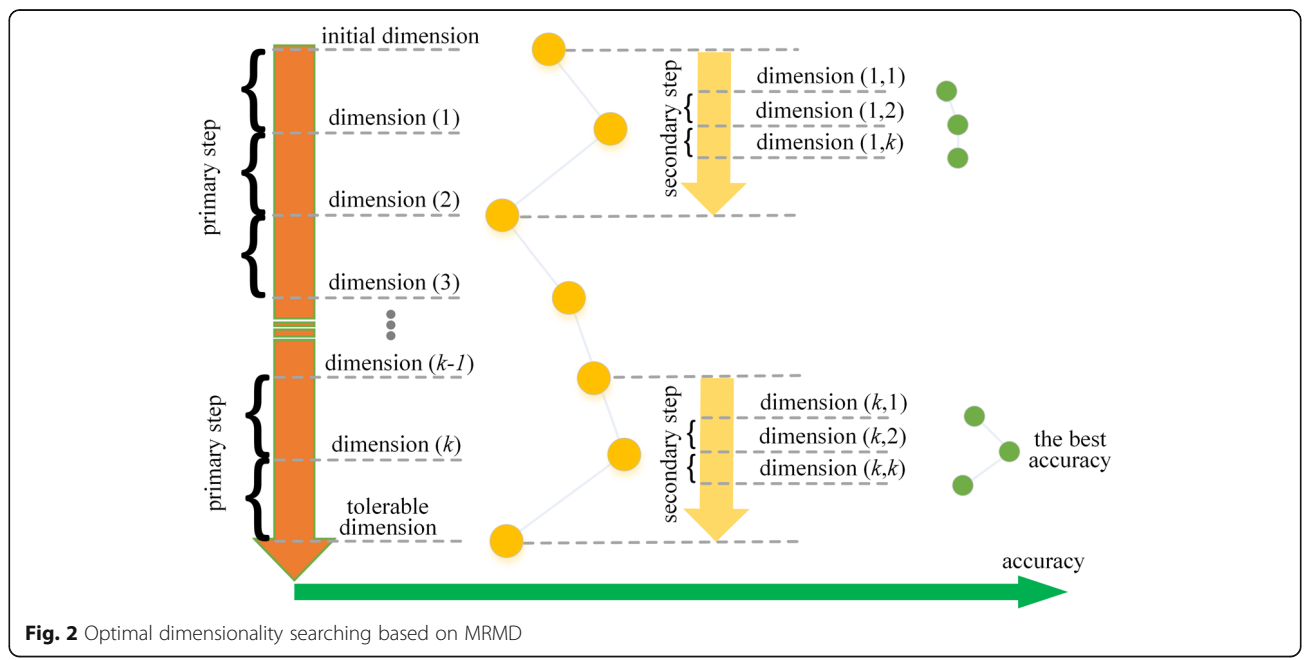

**Fig. 2** Optimal dimensionality searching based on MRMD

### Negative samples collection

There is no special database for the TBP negative sample, which is often appeared for other special protein identification problem. Here we constructed this negative dataset as followed. First, we list all the PFAM for all the positive ones. Then we randomly selected one protein from the remaining PFAMs. Although one TBP may belongs to several PFAMs, the size of negative samples is still far more than the positive ones. In order to get a high quality negative training set, we updated the negative training samples repeatly. First, we randomly select some negative proteins for training. Then the remaining negative proteins were predicted with the training model. Anyone who was predicted as positive was considered near to the classification boundary. These ones who had been misclassified would be updated to the training set and replace the former negative training samples. The process repeated several times unless the classification accuracy would not improve. The last negative training samples were selected for the prediction model.

The raw TBP dataset is downloaded from the Uniport database [34]. The dataset contains 964 TBP protein sequences. We clustered the raw dataset using CD-HIT [35] before each analysis, because of extensive redundancy in the raw data (including many repeat sequences). We found 559 positive instances (denoted $\Omega_{\text{Tata}}$) and 8465 negative instances at a clustering threshold value of 90%. Then 559 negative control sequences (denoted $\Omega_{\text{non} - \text{Tata}}$) were selected by random sampling from the 8465 sequence negative instances.

### Support vector machine (SVM)

Comparing with several classifiers, including random forest, KNN, C4.5, libD3C, we choose SVM as the classifier due to its best performance [36]. It can avoid the over-fitting problem and is suitable for the less sample problem [37–40].

The LIBSVM package [41, 42] was used in our study to implement SVM. The radial basis function (RBF) is chosen as the kernel function [43], and the parameter g is set as 0.5 and c is set as 128 according to the grid optimization.

We also tried the ensemble learning for imbalanced bioinformatics classification. However, the performance is as good as SVM while the running time is much more than SVM.

## Results

### Measurements

A series of experiments were performed to confirm the innovativeness and effectiveness of our method. First, we analyzed the effectiveness of extracted feature vectors based on pseudo amino acid composition and secondary structure, and compared this to 188D, PSIPRED, and 661D. Second, we showed the performance of our optimal dimensionality search under high dimensions, and compared these findings with the performance of an ensemble classifier. Finally, we estimated high quality negative sequences using an SVM, to multiply repeat the classification analysis.

Two important measures were used to assess the performance of individual classes: sensitivity(SN) and specificity(SP):

$$\text{SN} = \frac{\text{TP}}{\text{TP} + \text{FN}} \times 100\%$$

$$\text{SP} = \frac{\text{TP}}{\text{TN} + \text{FP}} \times 100\%$$

Additionally, overall accuracy (ACC) is defined as the ratio of correctly predicted samples over all tested samples [44, 45],:

$$\text{ACC} = \frac{\text{TP} + \text{TN}}{\text{TP} + \text{TN} + \text{FP} + \text{FN}} \times 100\%$$

where TP, TN, FN, and FP are the number of true positives, true negatives, false negatives, and false positives, respectively.

### Joint features outperform the single ones

We extracted composition and physicochemical fetures (188D), secondary structured features (473D), and the joint features (611D) for comparison. These data were trained, and the results of our 10-fold cross-validation were analyzed using Weka (version 3.7.13) [46]. We then calculated the SN, SP, and ACC values of five common and latest classifiers and illustrated the results in Figs. 3, 4, and 5.

We picked five different types of classifiers, with the aim of reflecting experimental accuracy more comprehensively. In turn: LibD3C is an ensemble classifier developed by Lin et al. [47]. LIBSVM is a simple support vector machine tool for classification developed by Chang et al. [41]. IBK [48] is a k-Nearest neighbors, non-parametric algorithm used for classification and regression. Random Forest [49, 50] is an implementation of a general random decision forest technique, and is an ensemble learning method for classification, regression, and other tasks. Bagging is a machine learning ensemble meta-algorithm designed to improve the stability and accuracy of machine learning algorithms used in statistical classification and regression. Using these five different category classification tests, we concluded that the combination of the composition-physicochemical features (188D) and the secondary structured features (473D) together is significantly superior to any single method, judging by ACC, SN, and SP values. In other words, neither physicochemical properties, nor secondary structure measurements alone can sufficiently reflect the functional characteristics of protein sequences enough to allow accurate prediction of protein sequence classification. A comprehensive consideration of both physicochemical properties and secondary structure can adequately reflect protein sequence functional characteristics. As for the type of classifier, LIBSVM had the best classification accuracy with our data, achieving up to 90.46% ACC with the 611D dataset. Furthermore, LIBSVM had better SN and SP indicator results than the other classifiers tested as well. These conclusions supported our consequent efforts to improve the current experiment using SVM, with hopes that we can obtain better performance while handling imbalanced datasets. Experiment in 4.3 will verify the SVM, but first we needed to consider another important issue. That is: what is the best dimensionality search method for reducing the 661D features dynamically to obtain a lower overall dimensionality and, thus, a higher accuracy with its final results.

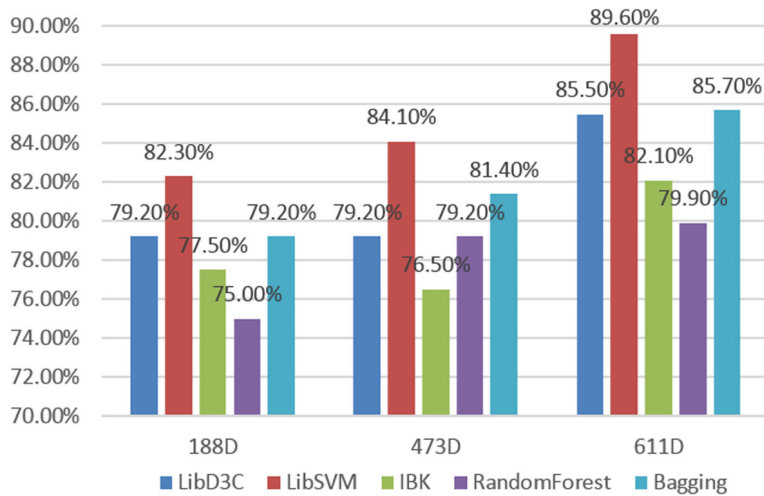

**Fig. 3** Five classifier sensitivities (SN)

### Dimensionality reduction outperforms the joint features

According to the former experiments, we concluded that the classification performance of 611D is far better than composition-physicochemical fetures (188D) or the secondary structured features (473D) alone, and that LIBSVM is the best classifier for our purposes. Then we tried MRMD to reduce the features. In order to save the estimating time, we first reduced 20 features every time, and compared the SN, SP, and ACC values, as shown in Fig. 6. We found that it performed better with 230–330 features. In the second step, we tried the features size with decreasing 2 every times from 230 to 330, as shown in Fig. 7. Optimal SN, SP, and ACC values are shown in Figs. 6 and 7 for each step.

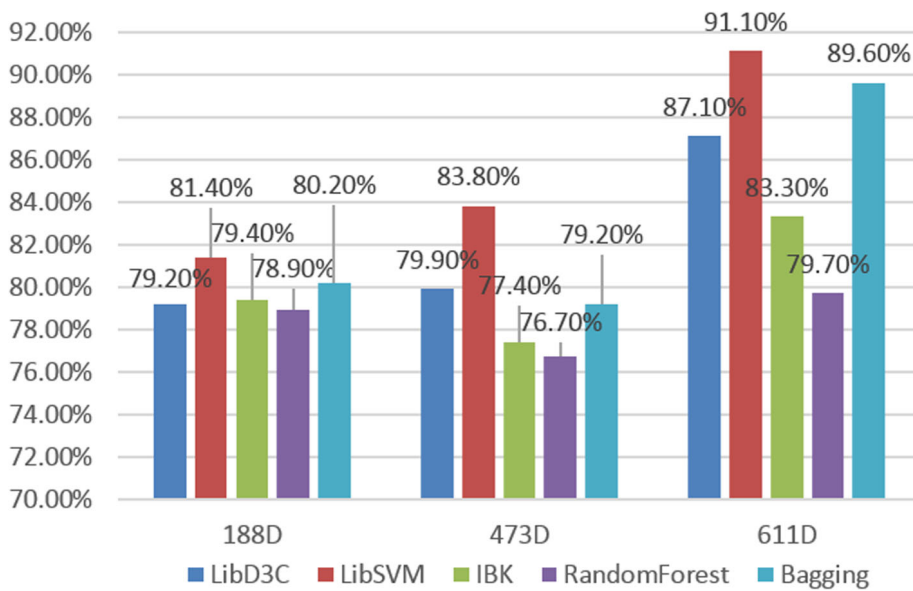

**Fig. 4** Five classifier specificities (SP)

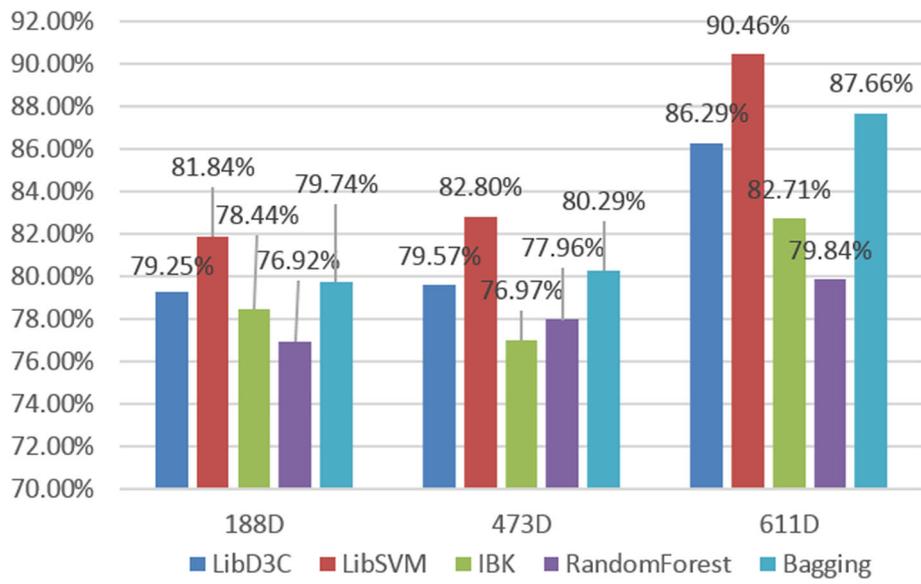

Fig. 5 Five classifier accuracies (ACC)

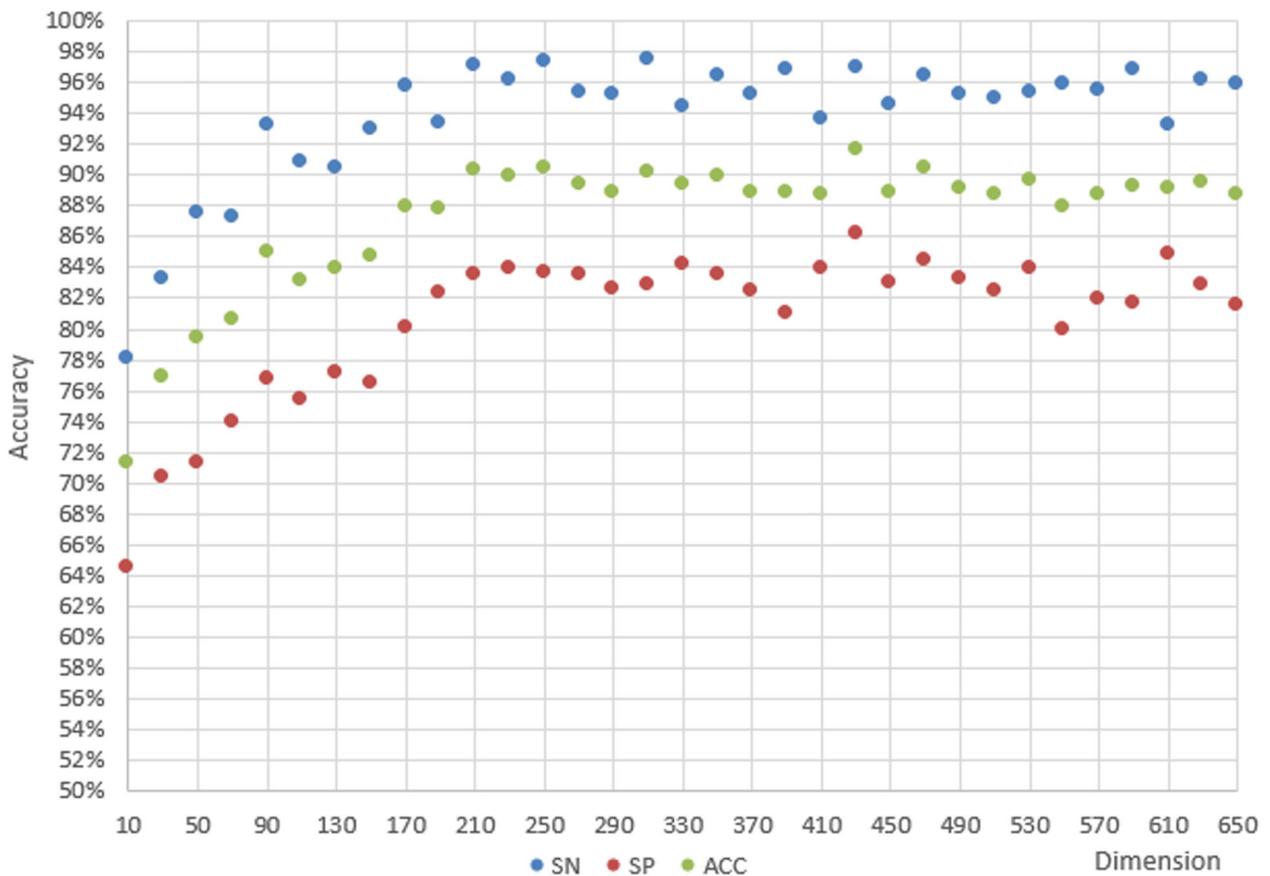

Fig. 6 SN, SP, and ACC of the primary step

The coarse search is illustrated in Fig. 6. The best ACC we obtained using LIBSVM is 91.58%, which is better than in the joint features in 4.1. Furthermore, ACC, SN, and SP all display outstanding results with combined optimal dimensionalities ranging from 220D to 330D. Figure 7 illustrates the elaborate search. The scatter plot displays the best ACC, SN, and SP values, 92.92, 98.60, and 87.30%, respectively. The scatter plot distribution suggested that there was no clear mathematical relationship between dimensionality and accuracy. Therefore, we considered whether our random selection algorithm is adequate to obtain the negative sequences in our dataset. We had to perform another experiment concerning the manner which we were obtaining our negative sequences to answer this question. We designed the next experiment to address the issue.

### Negative samples have been highly representative

In the previous experiments we selected randomly the negative dataset $\Omega_{non-Tata}$. It may be doubted that whether the random selection negative samples were representive and reconstruction of training dataset can improve the performance. Indeed, the positive and negative training samples are filtered with CD-HIT, which guaranteed the high diversity. Now we try to improve the quality of the negative samples and check whether the performance could be improved. We selected the negative samples randomly several times, and built the SVM classification models. Every time, we kept the support vectors negative samples. Then the support vectors negative samples construct a new high quality negative set, called plus $\Omega_{\text{non-Tata}}$.

The dataset is still $\Omega_{Tata}$ and plus $\Omega_{non-Tata}$, but now includes 559 positive sequences and 7908 negative sequences. First, the program extracts negative sequences from $\Omega_{Tata}$. The 20% of the original dataset that has the longest Euclidean distance will be reserved, and then the remaining 80% needed will be extracted from $\Omega_{non-Tata}$. Processing will not stop until the remaining negative sequences cannot supply $\Omega_{Tata}$. This process creates the highest quality negative dataset possible.

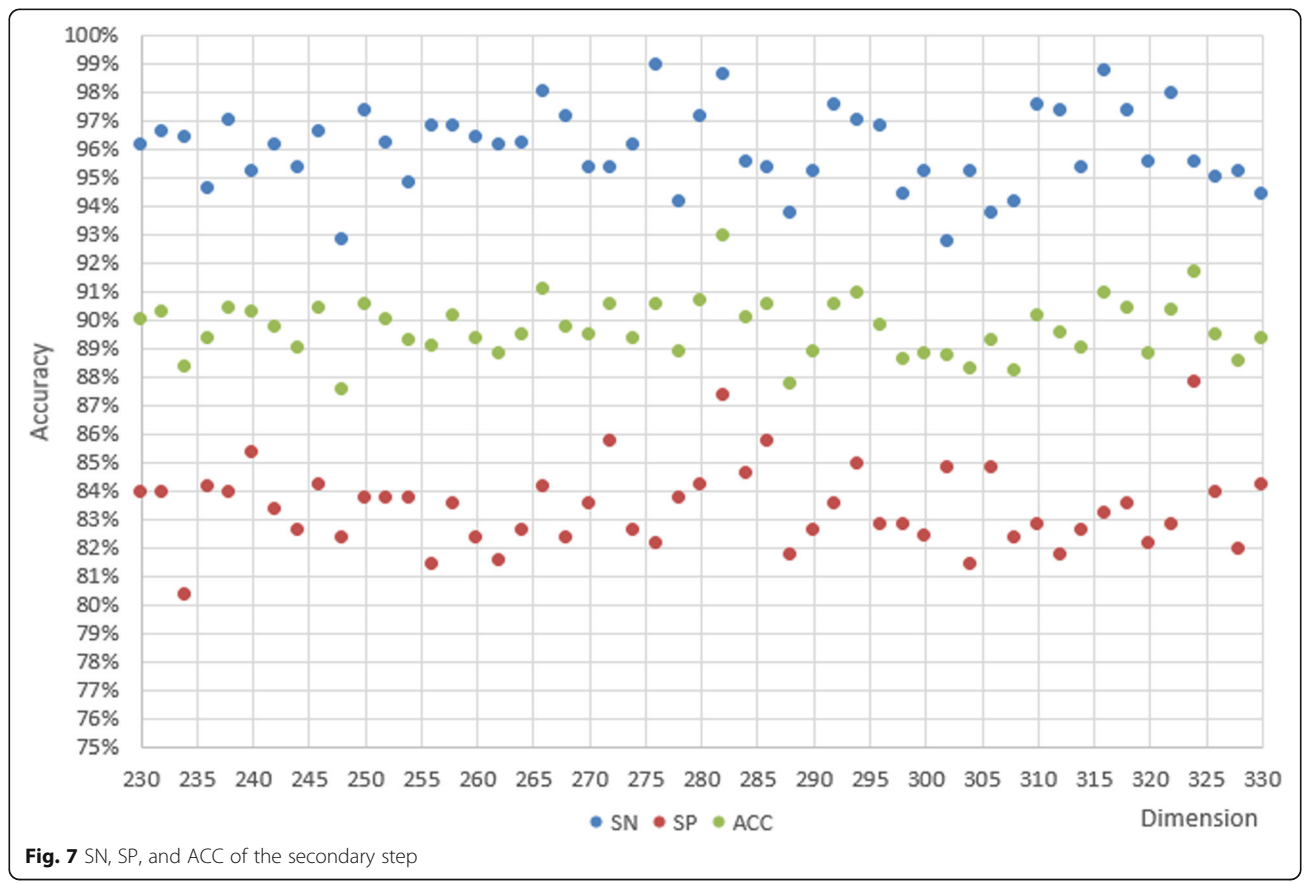

**Fig. 7** SN, SP, and ACC of the secondary step

Experiment in section 4.2 is then repeated with this highest quality negative dataset, instead of the random sample. Primary and secondary step values were estimated and PreTata was run to generate a scatter diagram illustrating dimensionality and accuracy.

Figure 8 illustrates the coarse search. The best ACC is 83.60% by LIBSVM at 350D. ACC, SN, and SP also show outstanding results with dimensionality ranging from 450D to 530D. Figures 9 illustrates the elaborate search. The scatter plot clearly displays the best ACC, SN, and SP as 84.05, 88.90, and 79.20%, respectively. However, we found that there was still no clear mathematical relationship between dimensionality and accuracy from this scatter plot distribution, and the performance of the experiment was no better than Experiment in section 4.2. In fact, the results may be even more misleading. We concluded that the negative sequences of experiment in section 4.2 were sufficiently equally distributed and had large enough differences between themselves. Although we selected high quality negative sequences with SVM in this experiment, the performance of classification and prediction did not improve. Furthermore, the ACC does not get higher and higher as dimensionality gets larger and larger, which is a characteristic of imbalanced data.

### Comparing with state-of-arts software tools

Since there is no TBP identification web server or tool with machine learning strategies to our knowledge, we can only test BLASTP and PSI-BLAST for TBP identification. We set $P$-value for BLASTP and PSI-BLAST less than 1. And the sequences with least $P$-value were selected. If it is a TBP sequence, we consider the query one as TBP; otherwise, the query protein is considered as non-TBP one. Sometimes, BLASTP or PSI-BLAST cannot output any result for some queries, where we record as a wrong result. Table 1 shows the SN, SP and ACC comparison. From Table 1 we can see that our method can outperform BLASTP and PSI-BLAST. Furthermore, for the no result queries in PSI-BLAST, our method can also predict well, which suggested that our method is also beneficial supplement to the searching tools.

### Discussion

With the rapidly increasing research datasets associated with NGS, an automatic platform with high prediction accuracy and efficiency is urgently needed. PreTata is pioneering work that can very quickly classify and predict TBPs from imbalanced datasets. Continuous improvement of our proposed method should facilitate even further researcher on theoretical prediction.

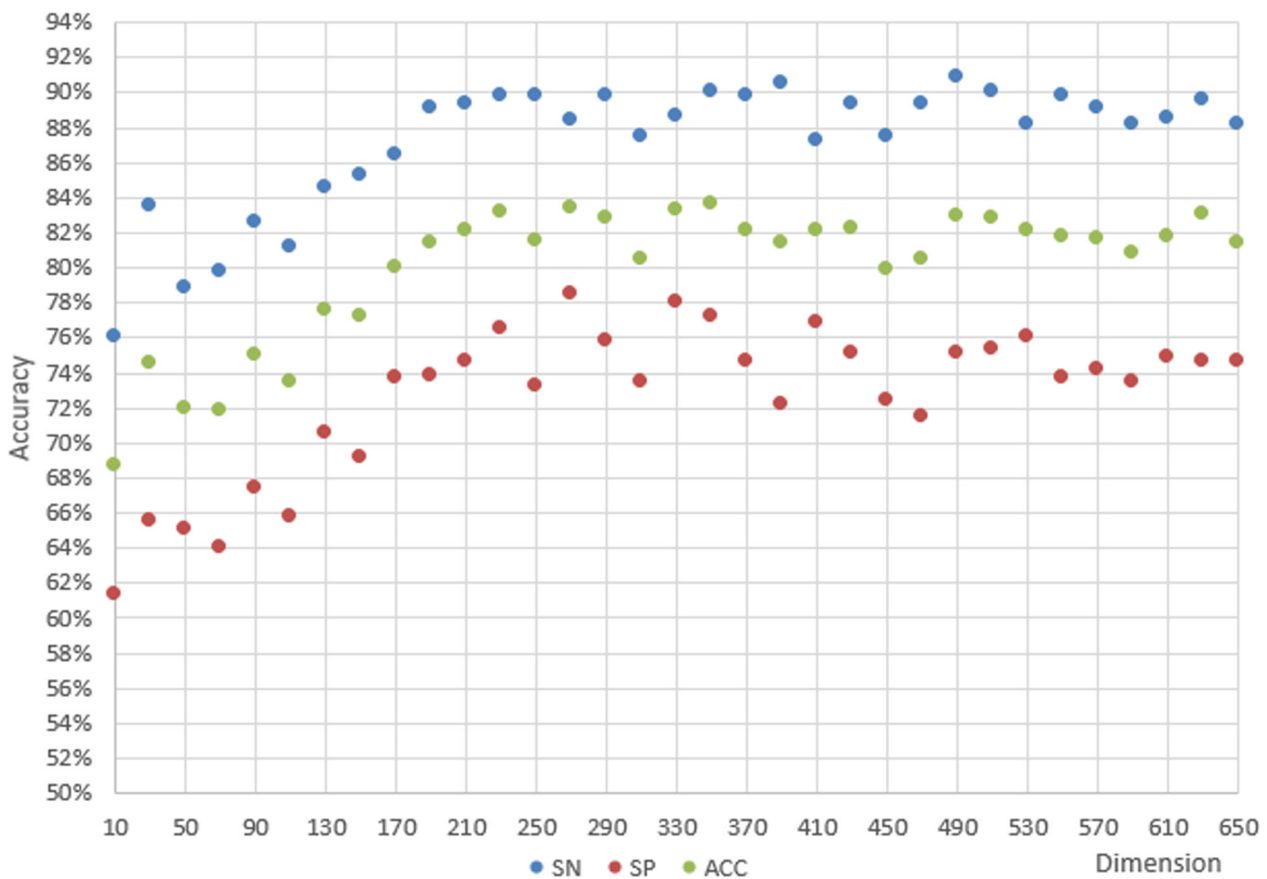

**Fig. 8** SN, SP, and ACC of the primary step with high quality negative samples

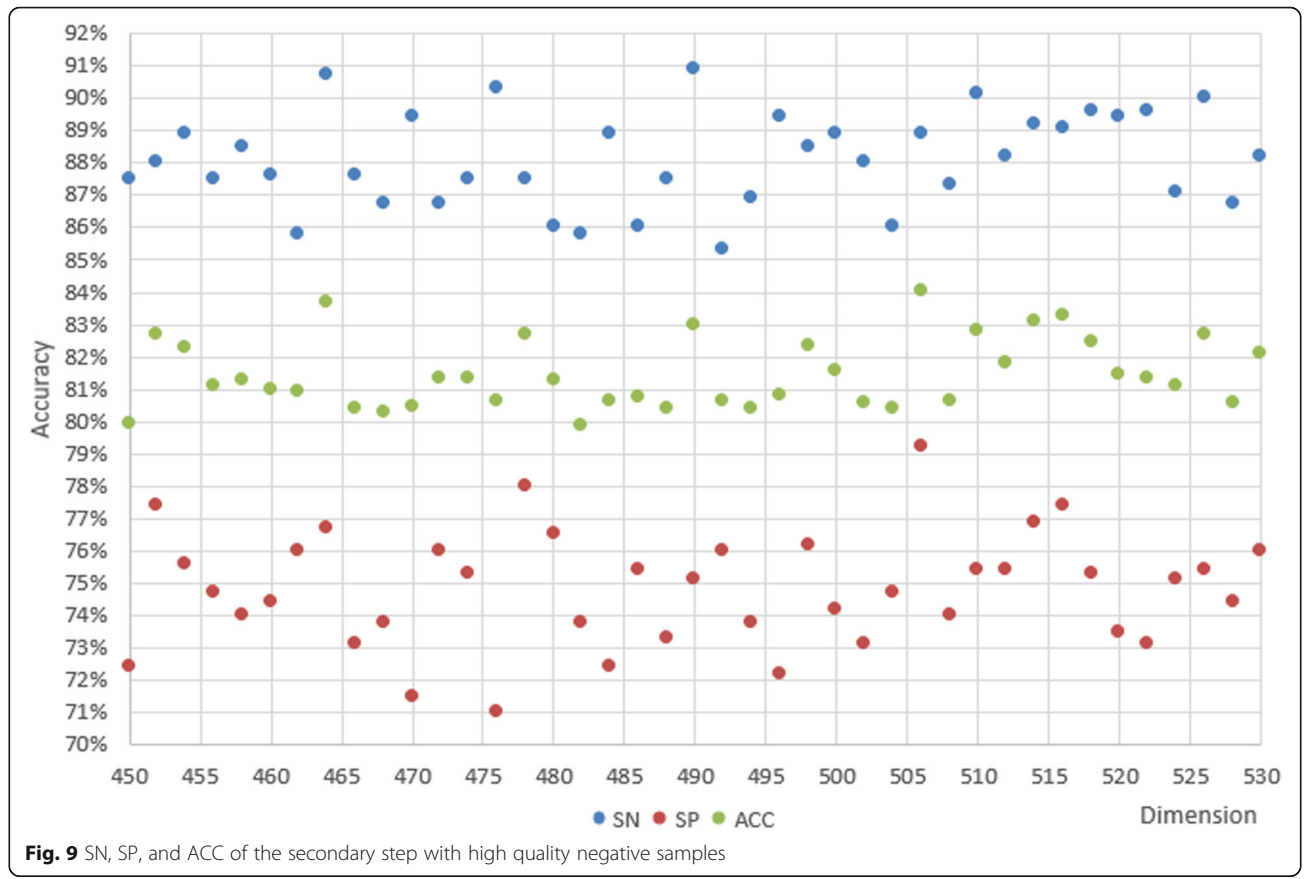

Fig. 9 SN, SP, and ACC of the secondary step with high quality negative samples

Our works employed advanced machine learning techniques and proposed novel protein sequence fingerprint features, which do not only facilitate TBP identification, but also guide for the other special protein detection from primary sequences.

## Conclusions

In this paper, we aimed at TBP identification with proper machine learning techniques. Three feature extraction methods are described: 188D based on physicochemical properties, 473D from PSIPRED secondary structure prediction results. Most importantly, we developed and describe PreTata, which is based on a secondary dimensionality search, and achieves better accuracy than other methods. The performance of our classification strategy and predictor demonstrates that our method is feasible and greatly improves prediction efficiency, thus allowing large-scale NGS data prediction to be practical. An online Web server and open source software that supports massive data processing were developed to facilitate our method's use. Our project can be freely accessed at http://server.malab.cn/preTata/. Currently, our method exceeds 90% accuracy in TBP prediction. A series of experiments demonstrated the effectiveness of our method.

Table 1 Comparison with the searching tools

|  | sn | sp | acc |
| --- | --- | --- | --- |
| Our method | 89.60% | 91.10% | 90.46% |
| BLASTP | 86.26% | 78.96% | 82.89% |
| PSI-BLAST | 88.62% | 81.60% | 84.36% |

**Abbreviations**
SVM: Support vector machine; TBP: TATA-binding protein

**Declarations**
This article has been published as part of BMC Systems Biology Volume 10, Supplement 4, 2016: Proceedings of the 27th International Conference on Genome Informatics: systems biology. The full contents of the supplement are available online at http://bmcsystbiol.biomedcentral.com/articles/supplements/volume-10-supplement-4.

**Funding**
This work and the publication costs were funded by the Natural Science Foundation of China (No. 61370010).

**Author details**
[1]School of Computer Science and Technology, Tianjin University, Tianjin, China. [2]Key Laboratory of Network Oriented Intelligent Computation, Harbin Institute of Technology Shenzhen Graduate School, Shenzhen, Guangdong 518055, China. [3]School of Information Science and Engineering, Xiamen University, Xiamen, China. [4]School of Computational Science and Engineering, University of South Carolina, Columbia, USA.

Published: 23 December 2016